# PanGeneHome : A Web Interface to Analyze Microbial Pangenomes

Camille Loiseau#, Victor Hatte#, Charlotte Andrieu, Loic Barlet, Audric Cologne, Romain De Oliveira, Lionel Ferrato-Berberian, Hélène Gardon, Damien Lauber, Mélanie Molinier, Stéphanie Monnerie, Kissi N'Gou, Benjamin Penaud, Olivier Pereira, Justine Picarle, Amandine Septier, Antoine Mahul, Jean-Christophe Charvy, and François Enault*

*Department of Biology, Clermont Auvergne University, Clermont-Ferrand, F-63000, France*

## Abstract

PanGeneHome is a web server dedicated to the analysis of available microbial pangenomes. For any prokaryotic taxon with at least three sequenced genomes, PanGeneHome provides (i) conservation level of genes, (ii) pangenome and core-genome curves, estimated pangenome size and other metrics, (iii) dendrograms based on gene content and average amino acid identity (AAI) for these genomes, and (iv) functional categories and metabolic pathways represented in the core, accessory and unique gene pools of the selected taxon. In addition, the results for these different analyses can be compared for any set of taxa. With the availability of 615 taxa, covering 182 species and 49 orders, PanGeneHome provides an easy way to get a glimpse on the pangenome of a microbial group of interest. The server and its documentation are available at http://pangenehome.lmge.uca.fr.

## Introduction

Recent advances in DNA sequencing technology led a rapid accumulation of microbial genomic data. Due to their inherent genomic plasticity, microbial species are now described throughout their pangenome and not just through a unique reference genome ([1–4]; for a review, see [5]). A microbial pangenome is composed of core genes (shared by all strains), dispensable/accessory genes (conserved in two or more strains) and genes unique to single strains. By comparing the number and conservation of genes across multiple genomes, researchers can thus gain insights into the genomic diversity, dynamics and evolution of a microbial taxon. Furthermore, the functional annotations of the core, accessory and unique genes can also be informative.

Several standalone tools (e.g. PGAP [6], PANNOTATOR [7], PanGP [8], Roary [9] and BPGA [10]) and web servers (e.g. Panseq [11], PGAT [12] and PanWeb [13]) dedicated to pangenome analysis have been developed recently and offer the possibility to compute pangenome analysis for genomes provided by a user (for a review of the different existing tools, see [14]). Among these, PGAT [12] is the only web site offering pre-computed pangenome analysis but only nine genus (representing 244 genomes) are available. Thus, as collecting genomes and running these tools implies a significant effort on the user side, we developed PanGeneHome, a Web server that offers precomputed pangenome analysis at a large scale for already sequenced genomes. Browsing PanGeneHome, pangenome analysis results can be directly accessed for any taxonomic level in the bacterial and archaeal trees for which the collection of publicly available genomes is sufficient (> 2 genomes).

## Methods

PanGeneHome provides users with a suite of tools for analyzing the pangenome of a selected taxon. To this end, the 2,674 bacterial and 167 archaeal complete genomes available in KEGG [15] were processed (Jan 2015). For any taxon (node or leaf) of the bacterial and archaeal trees that include at least three genomes, the pangenome was determined using KEGG Orthologous Clusters (OCs). These OCs were constructed using the 8,912,641 protein coding genes in all complete genomes (~ 3 % of the genes are encoded in plasmids). The KEGG functional categories to which these OCs are affiliated were included. KEGG orthologous (KO) groups [15], was also used to determine distances based on gene content and on the AAI of the shared genes.

The first step for the user is to select a taxon of interest, at any taxonomic level (phylum, class, order, family, genus or species level). This selection is made through a browsable and searchable tree, encoded using the jQuery plugin jsTree (https://www.jstree.com/). The different PanGeneHome sections described below were constructed as flat-file databases.

### Gene Conservation

For each taxon, the gene conservation was determined in two ways: (i) the conservation of all OCs of the taxon and (ii) the conservation of OCs in an average genome of this taxon. For a given taxon, the conservation of an OC is the percentage of genomes of this taxon in which this OC appears. These percentages were subsequently used to compute the distribution of the gene conservation for each genome, and these distributions were averaged for all the genomes of the considered taxon. These two results are displayed through interactive histograms with the Highcharts JavaScript library (http://www.highcharts.com).

### Pan- and Core-Genome Curves

The pan- and core-genome curves display respectively the total number of different OCs and the number of conserved OCs when considering an increasing number of genomes. When the number of possible genome combination was too large (e.g. there is more than 17 thousand billion different combination of 10 genomes out of 100), a random subset of 1,000 combinations was used. For each genome number considered, the average pan- and core-genome numbers is plotted with the standard deviation being represented by shaded zones around these two curves. Here again, Highcharts library (www.highcharts.com) was used to display these curves.

In addition, pangenome size, closedness and diversity were estimated using the R package micropan [16]. The Chao method was used to estimate the pan-genome size, a method that gives « *a conservative estimate, i.e. it tends to be on the smaller side of the true size* » [16]. To predict if a pangenome is open or closed, a Heaps law type of model was used and if the alpha value is below one, the pangenome is considered as open (see [17] for details). Finally, the genomic fluidity [18] was computed to quantify the pangenome diversity.

### Gene Content Tree

The distance between two genomes was defined as the fraction of genes unique to one of the two strains [19]. In details, for two genomes A and B, we counted the number of genes of A not present in B divided by the total number of genes of A. The dendrogram for each taxon was then determined by applying the PHYLIP Neighbor Joining method [20] to the corresponding distance matrix. To be able to compare evolutionarily distant species, KEGG KOs and not OCs were here considered. The tree for all bacteria is not available as it contains too many genomes (2674) to be visualized. These trees are displayed as circular and linear trees using the jsPhyloSVG JavaScript plugin [21] and full species name can be obtained by mousing over the corresponding leaf.

### AAI

The AAI of the shared genes between all genome pairs was determined as in [22]. Shortly, the identity percentage of all proteins inside a KO were determined using BLASTp [23], and for all genome pairs, the average amino acid identity was computed using all their shared proteins. The AAI between a pair of genomes A and B is thus the average of the amino acid identity using all the pairs of proteins of A and B that are present in the same orthologous group. For each taxon, the resulting AAI matrix was then used to build a dendrogram with the neighbor-joining method (PHYLIP suite [20]). Here again, KEGG KOs were used to enable the comparison of evolutionarily distant species, and dendrograms are displayed using the jsPhyloSVG plugin [21].

### Functional Analysis of Core, Accessory and Unique Genes

The functional annotations of the core, accessory and unique genes, defined here by the OC clustering, can also be displayed and compared. To this end, the KEGG pathway database was used [15]. First, the number and percentages of genes involved in the main categories (e.g. « Metabolism », « Genetic information processing », etc...) of this database were calculated for core, accessory and unique genes and displayed as histograms. Second, similar results were computed for a lower level of the pathway database (e.g. « Carbohydrate metabolism », « Replication and repair », etc.) through curves. These interactive histograms and curves were developed using the Highcharts JavaScript library (www.highcharts.com).

### Pan- and Core-Genome Comparison

Multiple taxa can be selected through an interactive tree, and the corresponding pan- and core-genome curves (defined previously in section 2) are displayed. The different metrics computed are also provided in a table.

### Core Function Comparison

Here again, multiple taxa can selected through an interactive tree, and the functional annotations of the core genes of each taxon are displayed through Highcharts histograms.

## Results and Discussion

PanGeneHome is a web server where pangenome analyses are available for all taxon that contain at least three sequenced genomes. The 2,841 genomes of KEGG allowed us to determine the pangenome of 10 phyla, 16 classes, 49 orders, 112 families, 164 genera and 182 species. Among these, 100 and 50 genera have respectively at least 5 and 10 sequenced genomes. This is to our knowledge the first web server where pangenomes are processed for all available genomes and for any taxonomic level.

### Details on the Protein Clustering Methods

The identification of orthologs is an important cornerstone for pangenome analysis. Here, two different clustering methods available in KEGG were used. The main difference between these methods is the granularity of the orthologous groups they produce :

- The OCs are constructed by automatically clustering proteins based on their sequence similarities and using a quasi-clique-based method [24]. All protein coding genes are thus included in the clustering and it produces fine-grained clusters. Indeed, the 8,912,641 KEGG proteins are clustered into 358,067 OCs (295 OCs are larger than 1,000 proteins) and 474,400 proteins remained as singletons.

- Complementary to this clustering into OCs, KEGG proteins are also assigned to KOs based on cross-species genome comparison using the KOALA (KEGG Orthology and Links Annotation) system [15]. The KOs produced do not include all proteins and are much larger than OCs: the 4,381,566 proteins assigned to a KO are clustered into 8,252 different KOs. As a comparison, the same proteins are clustered into 160,669 different OCs.

Contrary to KOs, OCs include all genes and were thus used to determine the different categories of a pangenome in a precise manner (core, accessory and unique genes). As KOs group even distantly related homologs (that are separated into several Ocs), KOs were used in gene content and AAI methods in order to compare evolutionarily distant genomes.

### Pangenehome through an Example

To illustrate the results that can be obtained with PanGeneHome, microbial species with various characteristics were here chosen. We selected three species described to have a closed pan-genome, namely *Bacillus anthracis* [17,25], *Buchnera aphidicola* [26] and *Campylobacter jejuni* [27], alongside three species that were described to have an open pangenome, namely *Bacillus thuringiensis* [28], *Propionibacterium acnes* [29] and *Prochlorococcus marinus* [17,30], and the species on which the concept of pangenome was initially tested, *Streptococcus agalactiae* [31].

The pangenome curves (Figure 1) and metrics (Table 1) are very different for these seven species considered. Indeed, the curves for *B. thuringiensis* and *P. marinus* keep increasing, even when more than 10 genomes are considered. Moreover, all metrics point out this trend as the estimated size of their pangenome are large (21,427 and 6,492 genes), their fluidity is larger than the one of the other species (> 0.18) and their Heap value lower (< 0.7). All this indicate that these species do have an open pangenome. Conversely, the pangenome curves of *B. anthracis* and *B. aphidicola* seems to

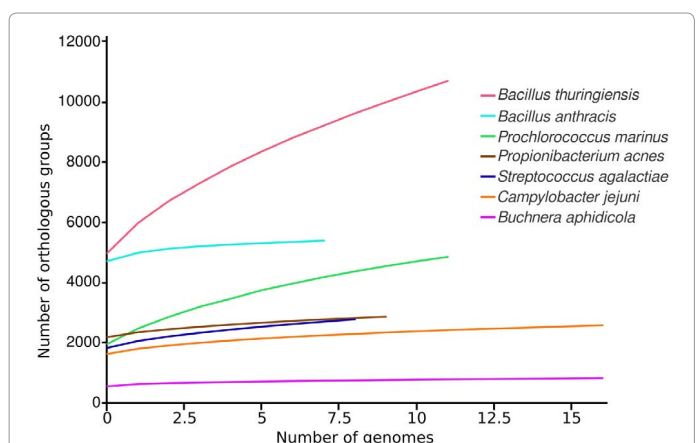

**Figure 1:** Pan-genome curves of seven different bacterial species. The average pan-genome size (i.e. the average number of total orthologous groups) is shown for the indicated number of genomes.

| Taxonomy | #Genomes | Chao | Heaps Alpha | Heaps Intercept | Fluidity Mean ± deviation |
|---|---|---|---|---|---|
| Bacillus thuringiensis | 12 | 21427 | 0.616 | 1499 | 0.188 ± 0.051 |
| Prochlorococcus marinus | 12 | 6492 | 0.644 | 792 | 0.272 ± 0.103 |
| Propionibacterium acnes | 10 | 3363 | 0.705 | 230 | 0.08 ± 0.023 |
| Campylobacter jejuni | 17 | 3426 | 0.812 | 280 | 0.11 ± 0.047 |
| Streptococcus agalactiae | 9 | 4036 | 0.842 | 414 | 0.137 ± 0.017 |
| Buchnera aphidicola | 17 | 1001 | 1.342 | 176 | 0.15 ± 0.124 |
| Bacillus anthracis | 8 | 5932 | 1.703 | 922 | 0.059 ± 0.023 |

Estimated pangenome size (Chao), closedness (Heaps Alpha and Intercept) and diversity (Genomic fluidity), computed using the R package micropan [16]. Taxon with a heaps alpha value below one are considered to have an open pangenome.

**Table 1:** Estimated pangenome metrics for different bacterial species.

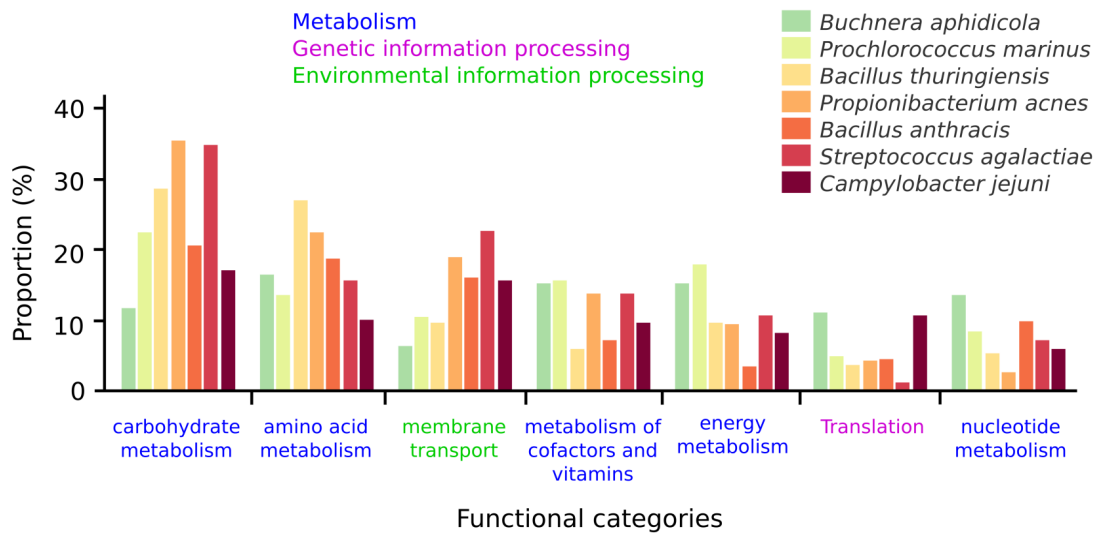

**Figure 2:** Functional categories of the core genes for seven bacterial species. The proportion of core genes involved in the different functional categories are displayed for each of the seven bacterial species selected. The categories are colored accordingly to the broader functional level to which they belong.

reach a plateau. Metrics also tend to show this trend, as their Heaps value are the only ones greater than one. The estimated pangenome of *B. anthracis* (5,932 genes) is this large because the individual genome of these bacteria are large (4,700 genes per genome), and its genomic fluidity is low (0.06). These two very different species in term of genome size and way of life (intracellular / free living) can be described as having a closed pangenome. Finally, the three last species considered here (*C. jejuni*, *P. acnes* and *S. agalactiae*) have similar trends : their pangenome curve have the same slope, their Heap value are lower than the 1 threshold (between 0.7 and 0.85) and the genomic fluidity is between 0.08 and 0.14. Thus, *C. jejuni*, described to have a closed pangenome [32], seems here quite similar to *S. agalactiae* and *P. acnes* that are described to have open pangenomes [17,29]. Moreover, these last two species do not present a pangenome as open as *B. thuringiensis* and *P. marinus* (Figure 1; Table 1). These results show the importance of using the same annotation and clustering methods to have comparable results, as the granularity of the clustering can have a dramatic impact on the pangenome estimate and metrics. It also highlights that each curve taken individually can lead to different interpretation, and comparing results for several species can provide additional information.

Another possibility offered by PanGeneHome is to compare the functional potential of core genes for the selected taxons, as in [33]. When considering the same 7 species, the core gene functions of *Buchnera aphidicola* are the most different to the ones of other species (Figure 2: average correlation of 0.67 between *B. aphidicola* core functional profiles and other species profiles). Indeed, nearly half of the annotated core genes of *Buchnera aphidicola* are involved in "Genetic information Processing", with 70 of the 157 core genes identified in the 17 genomes of this species being implied in translation. This result is not surprising as *B. aphidicola* is an endosymbiont of aphids that encode less than 600 genes and has lost lots of metabolic potential [34] such as anaerobic respiration, synthesis of phospholipids, complex carbohydrates, etc… The most similar species in terms of functional potential of their core genes are *P. acnes* and *S. agalactiae* (correlation of 0.93 for their functional profiles), two species having comparable pangenome closedness. More surprisingly, *Bacillus thuringiensis* and *Bacillus anthracis* are also similar in terms of functional potential of their core genes (correlation of 0.85) despite having opposite trends in terms of pangenome closedness. These two species belong to the Bacillus cereus group (NCBI taxonomy ID = 86661), and are actually thought to be part of the same species [35,36]. The AAI analysis of this "Bacillus cereus group" show that the sequenced strains of *B. anthracis* are more closely related to each other than the *B. thuringiensis* genomes (Figure 3). This last point might be due to the fact that *B. anthracis* strains are selected for culture and sequencing because of a precise phenotype (high toxicity). Selecting only very closely strains based on this phenoype might narrow the diversity of this group and artificially result in a closed pangenome. The fact that the core genes of these two species have similar functions reinforce the fact that the genomic characteristics of these two species might not be so different and that *B. anthracis* should be considered here as a subspecies.

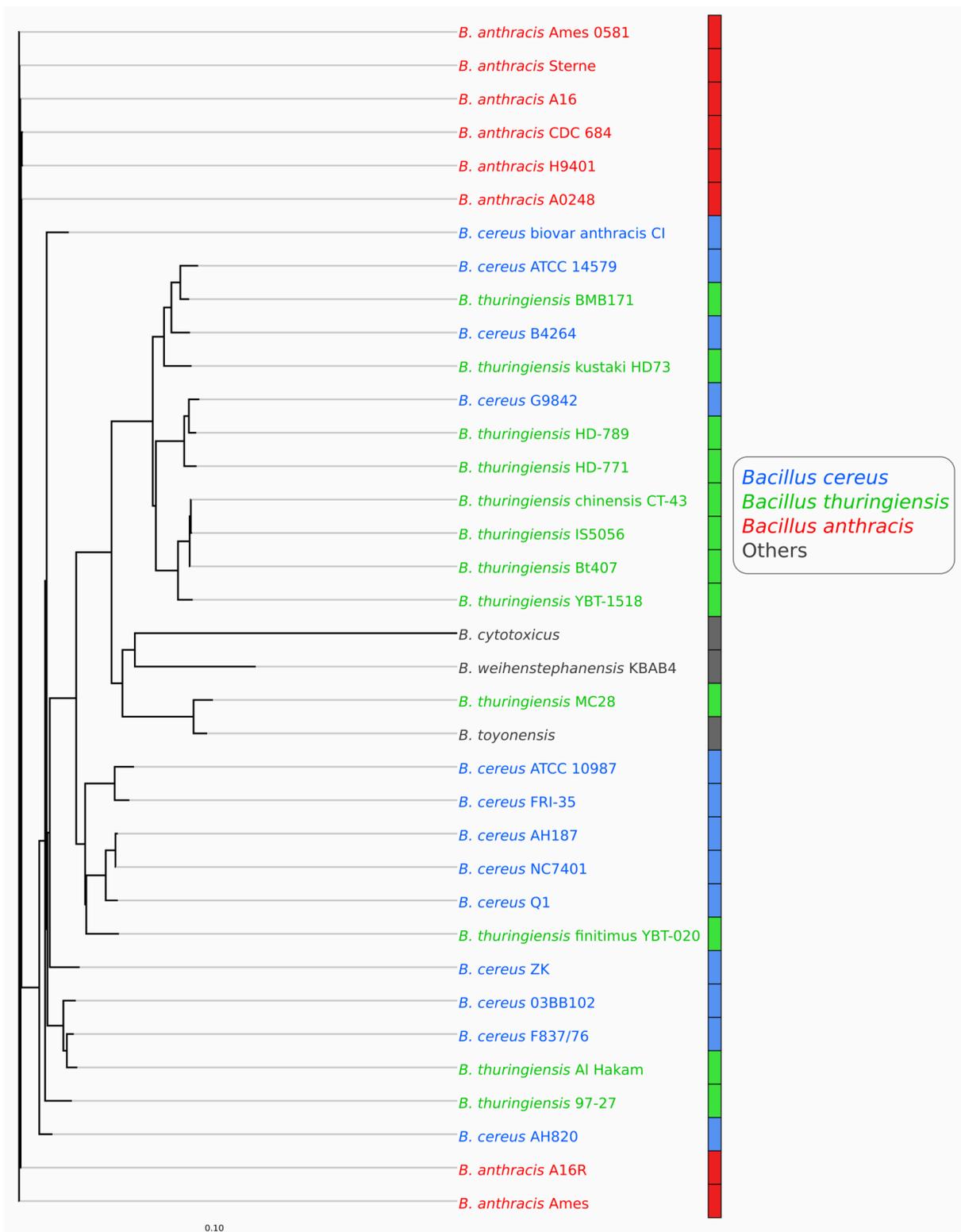

**Figure 3:** Dendrogram based on the AAI. Dendrogram based on the average amino acid identity of the shared genes (AAI) between all genome pairs of the Bacillus cereus group. The neighbor-joining method (PHYLIP suite [20]) was applied to the AAI matrix to build a dendrogram.

## Conclusion

A pangenome describes the full complement of genes in a clade or taxon. Even if pangenomes are typically analyzed at the species level, such analyses can be informative at any taxonomic level. PanGeneHome generates visualizations and metrics for pangenomes for all possible microbial clade, and as many as 615 taxa are available for analysis, including for example 182 different species and 49 different orders.

Pangenome metrics (size, diversity, closedness, etc...) are highly dependent on the genome annotation and protein clustering methods used. Pangenome studies often focus on one species and the results presented in separate studies are thus hardly comparable. Here, the annotation and clustering methods used were the same for all genomes, and pangenome results can be directly compared. Moreover, as highlighted by the results obtained for two Bacillus species, pangenome results should analyzed in regard of the diversity existing inside each taxon considered. Indeed, considering only evolutionarily close strains for a species will result in a low genomic fluidity and a closed pangenome, and analysis such as AAI should help in deciphering these evolutionary distances. Thus, PanGeneHome provides a comprehensive and uniform framework with a user-friendly interface to explore pangenomes for any microbial taxon, and should help microbiologists to quickly get a glimpse on the genomic plasticity and diversity for a clade of interest.

Considering the fast growing number of microbial genomes, the PanGeneHome tool will need to be updated regularly.

## Authors' Contributions

CL, VH, CA, LB, AC, RDO, LFB, HG, DL, MM, SM, KNG, BP, OP, JP and AS developed the Web site and CL and VH finalized its development. AM and JCC took care of the informatic infrastructure. FE conceived the study, coordinated the work and wrote the manuscript. All authors read and approved the final manuscript.

## Competing Interests

The authors have declared no competing interests.

## Acknowledgements

The authors thank Simon Roux for his careful reading of the manuscript. This research did not receive any specific grant from funding agencies in the public, commercial, or not-for-profit sectors.

**Corresponding author:** *François Enault, UMR CNRS 6023 Microorganisms: Genome and Environment, Build. A, 24 avenue des Landais, 63177 Aubière Cedex. La France, Tel: 33-(0)473- 407-471; Fax: 33-(0)473-407-670: E-mail:* francois.enault@uca.fr.

**Citation:** Loiseau C, Hatte V, Andrieu C, Barlet L, Cologne A, et al. (2017) PanGeneHome : A Web Interface to Analyze Microbial Pangenomes. J Bioinf Com Sys Bio 1(2): 108.